\documentclass[10pt,letterpaper]{article}
\usepackage[top=0.85in,left=2.75in,footskip=0.75in]{geometry}

\usepackage{amsmath,amssymb}

\usepackage{changepage}

\usepackage{textcomp,marvosym}

\usepackage{cite}

\usepackage{nameref,hyperref}

 \usepackage[right]{lineno}

\usepackage[nopatch=eqnum]{microtype}
\DisableLigatures[f]{encoding = *, family = * }

\usepackage[table]{xcolor}

\usepackage{array}

\newcolumntype{+}{!{\vrule width 2pt}}

\newlength\savedwidth

\newcommand\thickhline{\noalign{\global\savedwidth\arrayrulewidth\global\arrayrulewidth 2pt}%
\hline
\noalign{\global\arrayrulewidth\savedwidth}}

\raggedright
\usepackage[aboveskip=1pt,labelfont=bf,labelsep=period,justification=raggedright,singlelinecheck=off]{caption}

\makeatletter
\renewcommand{\@biblabel}[1]{\quad#1.}
\makeatother

\usepackage{lastpage,fancyhdr,graphicx}
\usepackage{epstopdf}
\fancyheadoffset[L]{2.25in}
\fancyfootoffset[L]{2.25in}
\begin{document}
\vspace*{0.2in}

\begin{flushleft}
{\Large
  \textbf{Elastic overtones: an equal temperament 12 tone music system with ``perfect'' fifths}
}
\newline
\\
Xavier Hernandez\textsuperscript{1*},
Luis Nasser\textsuperscript{2},
Pablo Garc\'{\i}a-Valenzuela\textsuperscript{3}
\\
\bigskip
\textbf{1}  Instituto de Astronom\'{\i}a, Universidad Nacional Aut\'{o}noma de M\'{e}xico, M\'exico City, M\'exico
\\
\textbf{2} Department of Physics, Loyola University Chicago, Chicago, Illinois, USA
\\
\textbf{3} School of Audio Engineering (SAE Institute Mexico), M\'exico City, M\'exico
\\
\bigskip

%
%





* xavier@astro.unam.mx

\end{flushleft}


%
\section*{Abstract}
The impossibility of a transposable 12 semitone tuning of the octave arises from the mathematical fact that $2 \times 2^{7/12} \neq 3$ i.e.,
the second harmonic of the fifth can not exactly match the third harmonic of the fundamental. This in turn, stems from the whole number harmonic structure
of western music, and the subsequent fundamental character of the octave interval as multiples of 2 in frequency, a property inherited by our music
system from the physics of instruments with vibrating elements being to a good approximation one dimensional. In the current era of electronic music, one can relax the
above assumptions to construct an analogous music system where all the structural properties of the standard music system are preserved, but where harmonics are
not whole number multiples of the fundamental frequency, and the octave is no longer a factor of 2 in frequency. This now allows to construct a transposable 12
semitone music system where the second harmonic of the fifth exactly matches the third harmonic of the fundamental. The enhanced harmonic qualities of this
system recover to a good approximation the musical qualities of Just Intonation, whilst retaining by construction all the versatility and modulating ability of
12TET.




%

\section*{Introduction}

The standard musical system of western music is composed of a set of 12 notes. Each note in turn, consists of a sound having a base frequency, to which we shall refer as the
1st harmonic of the note in question, and the addition of higher harmonics. The frequency of these higher harmonics are integer multiples of the base frequency of the note
in question. Thus, the 2nd harmonic has double the frequency of the base frequency, the third harmonic has three times the frequency of the base frequency, and so on. The set of
amplitudes of each of these components - the frequency spectra of a given note - yields what is termed the timber of the note in question, and is what distinguishes different
instruments, or different ways of playing an instrument, when sounding the same musical note. This structure is not so much an aesthetic or stylistic choice, but is in fact a
feature inherited from the physics of vibrating objects that are effectively close to one dimensional: the strings of a guitar, piano, violin or harp, or the air inside a flute,
are all to a first approximation, one dimensional systems. This dimensionality in turn implies that the objects in question will vibrate harmonically at a collection of frequencies
that, to a very good approximation, are integer multiples of a fundamental value.

This whole number multiple harmonic feature of notes has an important implication: it implies that a note which has a base frequency of double that of another one, will have as its
harmonics frequencies which are all included in the harmonics of the note of lower base frequency. In other words, the harmonic overtone series of a note of base frequency $f$ are:
\begin{equation}
\label{fharmonics}
 (f, 2f, 3f, 4f, 5f, 6f, 7f, 8f, 9f,10f,\cdots, \infty).
\end{equation}
Note that while the frequencies can mathematically increase to $\infty$, in practice this is not relevant because, asides from physical limitations cutting in at sufficiently
high frequencies, the largest frequency heard by the average human adult is $f_{Max}=20$ kHz. For a note of base frequency $2f$, the harmonic overtones are also integer multiples
of the base frequency: 
 \begin{equation}
\label{2fharmonics}
 (2f, 4f, 6f,8f, 10f,12f,\cdots,\infty),
\end{equation}
 and we can see from Eqs (\ref{fharmonics}) and (\ref{2fharmonics}) that every harmonic of $2f$ is contained in the harmonics of $f$, including the base note $2f$ itself.
 This complete overlap of harmonics is what gives rise to the perception that, despite their difference in pitch, a note of frequency $f$ and a note of frequency $2f$ are
 in essence the same note, and their pitch difference is what we call one octave. The octave is hence the primary structural division and most consonant interval in the
 standard western music system.

 Once the fundamental importance of the octave is established, the issue of dividing the octave arises; into exactly how many notes should the octave be divided, and how
 should these notes be spaced within it? While many variants on the division of the octave have been implemented in human history, we will focus on a solution that was
 first proposed by Chinese prince Zhu Zaiyu in 1584 and independently by Flemish engineer Simon Stevin in 1585 \cite{bib5}. This solution is called ``12 tone equal temperament'',
 or 12-TET for short, and it divides the octave into 12 equally spaced notes. For future reference, the interval between successive notes in 12-TET is called a semitone.
 12-TET was eventually adopted by Western composers in the 18th-19th centuries. Recently, it has been suggested e.g.,~\cite{bib1}, ~\cite{bib24}, that the minimization of a
 suitably constructed "musical free energy" - analogous to the Helmholtz free energy in thermodynamics - yields a 12 note division of the octave among others. What is clear
 is that if we wish to access one millennia of documented music in the Western music catalogue, any modified musical system used to perform this catalogue requires a 12 note
 octave. However, as we shall discuss later in more detail, such an alternative musical system need not be restricted by the requirement that an octave be an interval
 defined by a doubling of frequency. Indeed, the defining consonant element of the octave is only the requirement that every harmonic of the higher note be contained within
 the harmonics of the lower one, not that the higher pitch be double that of the lower one.

After the octave, the most consonant interval in Western music is the perfect fifth, defined as a frequency interval of 3:2 between the base frequencies
of two notes. Thus, the harmonic series of a note of base frequency $3f/2$ is:
\begin{equation}
\label{fifthharmonics}
\left(\frac{3f}{2},3f,\frac{9f}{2},6f,\frac{15f}{2}, 9f,\cdots,\infty \right).
\end{equation}

It is easy to check from Eqs (\ref{fifthharmonics}) and (\ref{fharmonics}) that all even harmonics of $3f/2$ are also harmonics of  $f$, i.e. half the harmonics of the perfect
fifth are harmonics of the fundamental. The fact that the perfect fifth has such a high degree of consonance with the root note has naturally prompted the
desire to use the perfect fifth as a key building block in the spacing of the 12 intervals within the octave. 

The first example of this is known as Pythagorean tuning, although its origins date back to Mesopotamia \cite{bib25}. The basic idea is to start with a fundamental frequency
and raise it by a fifth (multiply it by $3/2$). You will keep raising by perfect fifths until the number you obtain is larger than twice the fundamental frequency, at which
point you lower by an octave (divide by 2). You will do this to generate 6 notes; the other 6 are obtained by reducing the fundamental in perfect fifths (dividing by $3/2$)
and if the answer obtained is less that the fundamental frequency, you raise it by an octave (multiply by 2). For example, take a fundamental frequency $f$. You raise it by
a fifth to obtain to $3f/2$. Because $3f/2<2f$, we raise by a fifth again to $9f/4$, but this is now greater than $2f$, and so we divide by two to get $9f/8$. We raise by a
fifth again to get $27f/16$ which still lies within the octave. We raise this by a fifth to get $81f/32$, which is greater than $2f$ and we thus divide by 2. We continue this
process of raising by fifths, and dividing by 2 if the result is greater than $2f$ until we have the first six notes. We then go back to $f$, and divide by 2/3, which gives
us $2f/3$. Because this is less than $f$, we raise it by an octave to $4f/3$, and we do this to obtain the remaining 5 notes that divide the octave into 12 unevenly separated
notes. This method generates a major scale as shown in the table below:
\begin{table}[h!]
\caption{Pythagorean Tuning, Major Scale.}
\label{}
\begin{tabular}{lll}
\textbf{Interval} &\textbf{Frequency Ratio}  & \textbf{Formula}   \\
   Unison    & $1:1$ &  $3^{0}\times2^{0}$  \\
      Major Second & $9:8$ & $\left(\frac{3}{2}\right)^{2}\times\frac{1}{2}$   \\
      Major Third & $81:64$ &  $\left(\frac{3}{2}\right)^{4}\times\left(\frac{1}{2}\right)^{2}$  \\
       Perfect Fourth& $4:3$ & $\frac{2}{3}\times2$   \\
       Perfect  Fifth&$3:2$  & $3^{1}\times2^{-1}$   \\
     Major Sixth  &$27:16$  & $\left(\frac{3}{2}\right)^{3}\times\frac{1}{2}$   \\
       Major Seventh& $243:128$ &  $\left(\frac{3}{2}\right)^{5}\times\left(\frac{1}{2}\right)^{2}$   \\
       Octave& $2:1$ & $3^{0}\times2^{1}$      
\end{tabular}
\end{table}

This tuning system is perfectly fine when used to tune instruments such as the Greek tetrachords, which spanned one octave \cite{bib26}. However, if you wish to go into
extended octaves and polyphonic music there is a problem: using a 12 tone division of the octave, the perfect fifth is an interval
of 7 semitones. This means that 12 perfect fifths should equal 7 octaves. Upon calculation, we see that:
\begin{eqnarray}
\left(\frac{3}{2}\right)^{12 }& = & \frac{531441}{4096}\nonumber \\
& \approx & 129.746,\\
2^{7}&=&128\nonumber\\
\Rightarrow 2^{7/12}& \neq & 3/2.
\end{eqnarray}

Note that while $\delta$, defined as:
\begin{eqnarray}
\delta &=&\left |2^{7/12}-\frac{3}{2}\right |\nonumber\\
& = &|1.4938307\cdots-1.5| \nonumber \\
 & = & 0.001693\cdots
\end{eqnarray}
appears to be small enough to simply say $\delta\approx0$ is a good approximation, in practice this small discrepancy leads to an unacceptable dissonance called the
Pythagorean comma (aka the wolf tone) and thus the need for other solutions to the problem of the division of the octave for use in polyphonic music. 

How exactly should the 12 semitones within the octave be spaced? Just like having two notes separated by a frequency ratio of 3:2 leads to 50\% of shared harmonics, other
intervals have high degrees of consonance. An example of a tuning system that exploits all the perfect intervals of the Pythagorean system while introducing new, simpler
ones is called Just Intonation (JI) \cite{bib27}. In this system, two notes separated by a frequency ratio of 4:3 are still called a perfect fourth. We can see that the
harmonics of a perfect fourth are:
\begin{equation}
\label{fourth}
\left(\frac{4f}{3}, \frac{8f}{3}, 4f, \frac{16f}{3}, \frac{20f}{3}, 8f,\cdots,\infty \right)
\end{equation}

A comparison of Eqs (\ref{fharmonics}) and (\ref{fourth}) shows that a perfect fourth shares one third of its harmonics with $f$ and is the most consonant interval after the
perfect fifth. Another important interval in JI is the major third, defined as two notes separated by a frequency ratio of $5:4$ and whose harmonics are:
\begin{equation}
\label{majthird}
\left(\frac{5f}{4},\frac{10f}{4},\frac{15f}{4},5f,\frac{25f}{4},\frac{30f}{4},\frac{35f}{4},10f,\cdots,\infty \right),
\end{equation} 
This is much simpler than the major third of Pythagorean tuning where the frequencies are in the ratio $81:64$, and we can see from Eqs (\ref{fharmonics}) and (\ref{majthird})
that the JI major third shares 25\% of its harmonics with the fundamental. In JI, a major sixth interval has frequencies in the ratio $5:3$ which is simpler than the
Pythagorean $27:16$, and likewise a JI Major seventh has frequencies in the ratio $15:8$, in contrast to the Pythagorean frequency ratio of $243:128$. Consonant intervals
like these all share one thing in common: shared harmonics will occur whenever the frequency ratio of the notes is rational, and hence, tuning systems where these type of
intervals played a major role were initially popular. However, these schemes can not yield uniformly spaced semitones; if we want the fifth to correspond to 7 semitones,
under a constant spacing scheme with 12 semitones, we would require seven intervals, $2^{7/12}$ to be equal to $3/2$, which is simply not the case. 

By itself, JI gives the listener a highly consonant musical experience, but there is a very serious limitation. If the composer wishes to change keys (a practice known
as modulation), the intervals for the new key will be in conflict with the structure of the scale in JI, giving rise to dissonance and more ``wolf tones'' \cite{bib23}.
To see why this is so, consider a scale of C major in JI, which consists of the notes C,D,E,F,G,A and B before reaching the higher C and completing the octave. The frequencies
of the notes are:
\begin{equation}
\label{cmajorJI}
f_{C},\frac{9f_{C}}{8},\frac{5f_{C}}{4},\frac{4f_{C}}{3},\frac{3f_{C}}{2},\frac{5f_{C}}{3},\frac{15f_{C}}{8},2f_{C}.
\end{equation}
Now, suppose you wish to change key from C major to D major. The D major scale has notes D, E, F$\sharp$, G, A, B and C$\sharp$ before ending in the D one octave above the
starting note. If we look at the interval between the first and the second notes in C major, we simply obtain $9:8$. The interval between the first and the second notes in
D major should also be in the ratio $9:8$. However, when we calculate it we obtain:
\begin{eqnarray}
\frac{\left(\frac{5f_{C}}{4}\right)}{\left(\frac{9f_{C}}{8}\right)} & = & \frac{10}{9}\nonumber \\
 &\neq&\frac{9}{8}.\nonumber
\end{eqnarray}
Other pairs of successive notes in the scale of D major will likewise be dissonant with respect to their counterparts in the scale of C major, which means modulation is
not possible within JI. It was therefore necessary to find another way to divide the octave that would allow for modulation to happen freely and accurately. That division
of the octave is the afore-mentioned 12 TET. In this tuning system, the pythagorean predilection for rational frequency ratios is abandoned for all intervals except the
octave, which is still the span between a note of frequency $f$ and one with frequency $2f$. Everything else in-between is evenly spaced into equal semitones that differ
from the note preceding it by a factor $2^{1/12}$ such that 12 semitones span the octave. In 12 TET, the scale of C major is written as
 \begin{equation}
\label{C12tet}
f_{C},f_{C}\times2^{2/12},f_{C}\times2^{4/12},f_{C}\times2^{5/12},f_{C}\times2^{7/12},f_{C}\times2^{9/12},f_{C}\times2^{11/12}, 2f_{C}.
\end{equation}
 The loss of perfect intervals is replaced by the ability to modulate at will. In a sense, it is like seeing things with the eye: when you focus your vision on an object,
 there is one unique focal length for the optical lens. Everything which is not at that exact distance from the optical lens will, strictly speaking, be out of focus, but
 we are so used to the slight blur of everything else in the field of view that it does not hamper our eyesight. For this reason, 12TET is a powerful way to play polyphonic
 music in constantly shifting keys, and it is the standard tuning system used in western music. However, there is a valid issue concerning the timbre of the notes in 12TET
 versus those of Just intonation. After all, the ``blur'' may be small, but what it means is that no interval is properly tuned except for the octave. The overtones of 12TET
 notes beat against each other, and the result is an egalitarian bedrock of dissonance which can be problematic for listeners accustomed to perfect intervals.



\section*{Proposal of a perfect fifth$^{\star}$ music system that can modulate.}

As seen in the previous section, a music system which has octaves divided into 12 equal semitones cannot have perfect fifths. This is directly due to the choice of the
octave as the interval $(f,2f)$ and the mathematically inescapable reality that $2^{7/12} \neq 3/2$. However, when we think of the origin of the octave, we can see that
what makes that interval the foundational building block of harmony is the fact that all the harmonics of $2f$ are contained in the harmonics of $f$. The use of the factor
of 2 is neither an aesthetic choice nor a feature of the physiology of the human ear, but simply a consequence of the physics of instruments where the vibrating element
is to a good approximation one dimensional. In such instruments, sound produced is described using sinusoidal functions; the factor of 2 stems from the structure of the
solutions to the wave equation in one spatial dimension. Of course, many instruments exist such as drums, bells and cymbals that cannot be modeled as one-dimensional
vibrating systems. We know that the modes of vibration of such instruments are not described by sinusoidal functions, and are therefore anharmonic. Consequently, they
are slightly out of tune with other harmonic instruments, and yet they serve a useful - even ubiquitous - function in music. Now, if we focus on trying to construct a
different tuning system with a functional octave whilst regaining a perfect fifth, then from all we have seen it is clear that doubling the frequency to obtain the octave
consonance should be discarded as a strict necessity; as long as we have a system wherein all the harmonics of one note are contained in the harmonics of an other, we
have a functional octave which can be used to build a music system. Such a system would - by necessity - have to be implemented in electronic music, where
oscillations are not happening in what are effectively one-dimensional vibrating strings or columns of air.

Over the last three decades, the advent of electronic music has led to instruments where the mix of frequencies
produced when a given note is played are in fact specifically engineered \cite{bib28}. Most of these efforts have hinged on microtonality, and octaves divided in many more
notes than 12, which essentially renders them unable to coexist with the overwhelming majority of Western music. While a small number of tuning system have been proposed
where the choice of frequencies per note has been selected to produce variant musical systems with a 12 note octave, it is surprising that none have appeared exploring
beyond the standard rigid integer overtone structure. Examples of the former can be found in meantone temperament ideas, which emphasise pure major thirds at the expense of
flattening the fifths, but which are plagued by wolf tones and hence unusable when modulating away from the tonic \cite{bib30}. Limit Tuning proposals, which are a variation
on JI which creates different 12-note octaves for specific keys have also been tried, but by their very nature are not practical in a modern musical setting that requires tonal
versatility \cite{bib29}. We mention also the Fifth-Tone tuning ideas, developed by violinist Maria Renold in 1962 \cite{bib4}, \cite{bib7}, a non-tempered tuning system that
utilises perfect fifths and fourths centred on an $A_{4}$ with a frequency of 432 Hz instead of the customary 440 Hz. Unfortunately, while these type of tuning systems are often
praised for ``sounding better'', they face challenges when it comes to compatibility with standard equal temperament instruments.

In wanting to preserve the overall structure of western music, and hence retain a catalogue of close to two millennia, we find it desirable to retain the octave,
at least closely. Let us think of a music system where an analogue of the octave, the octave$^{\star}$ exists. Here, a fundamental note having a first harmonic with a frequency of
1, will be separated by an octave$^{\star}$ when the first harmonic of the octave$^{\star}$ occurs at a frequency ratio of $x$, with $x$ a real number to be optimally determined.
To preserve the analogy with the
standard musical system, we want $x\approx 2$. Also, all notes will have harmonics$^{\star}$ which appear at frequency ratios of $x, \frac{1}{2}(x+x^{2}), x^{2},
\frac{1}{4}(3x^{2}+x^{3}), \frac{1}{4}(2x^{2}+2x^{3}), \frac{1}{4}(1x^{2}+3x^{3}), x^{3}, \frac{1}{8}(7x^{3}+x^{4}),  \frac{1}{8}(6x^{3}+2x^{4})....$.  The  Nth harmonic will be given by:
\begin{equation}
  \frac{1}{M} \left[(M-L)x^{\alpha}+Lx^{\beta}\right],
\end{equation}

\noindent  where $M=Pp$, and $Pp$ is the previous power of 2 smaller than $N$ e.g., when $N=13$, the previous power of 2 is 8, so that $Pp=8$ and
$M=8$. Similarly, the next power of 2 with respect to $N$ is $Pn$, and $\alpha$ is given by the index of $Pp$ and $\beta$ is the index of $Pn$. $L$ is given by $N-Pp$. Thus,
for the 13th harmonic the previous power of 2 is 8, $Pp=8=2^{3}$, the next power of 2 is 16, $Pn=16=2^{4}$, making $M=8$, $\alpha=3$ and $\beta=4$. Since $L=N-Pp$,
$L=13-8=5$, and the 13th harmonic is $\frac{1}{8}(3x^{3}+5x^{4})$. In what follows, we shall refer to analogous entities in the proposed music system by the same name as used
in the standard music system, with a $\star$ index.

It is easy to verify that under such a scheme, the octave$^{\star}$ preserves the structure of the octave: all harmonics of any octave$^{\star}$ are present in all previous
octaves$^{\star}$, as shown in Table~\ref{table1}. Also, when $x=2$ we recover the standard musical system and the octave. Just as with the octave, with respect to the following
octave$^{\star}$, all even harmonics of any given note will also be included in the following octave$^{\star}$. We now want a transposable 12 semitone structure, so that the
octave$^{\star}$ will be divided into 12 equally spaced frequency intervals. The fundamental frequency of the notes in an Octave$^{\star}$ will be given by powers of $x^{1/12}$
times that of the first note in said Octave$^{\star}$.

Since the most important consonance in the perfect fifth is that the
second harmonic of the perfect fifth equals the third harmonic of the fundamental, we can demand that:

\begin{equation}
  x^{1+7/12} = \frac{1}{2}(x+x^{2}),
\end{equation}

\noindent dividing by x yields:

\begin{equation}
\label{xeq}
2 x^{7/12} = x + 1.
\end{equation}

Eq (\ref{xeq}) has two solutions: a trivial one when $x=1$, and all harmonics and notes have always the same reference frequency, wherever this frequency is chosen.
It is a maximally consonant music system consisting of only one frequency, an extremely uninteresting solution which we can term the "Ommmmmm" solution. However, a
more interesting solution exists which can be found numerically at $x=1.9725786270914....$. This solution is very close to the $x=2$ of the standard music system, and hence
an interesting candidate for an analogous music system where the fifth is close to perfect. In the perfect fifth of the standard music system, all even harmonics of the perfect
fifth occur also as harmonics of the fundamental. In the perfect$^{\star}$ system presented, this only happens for harmonics of the fifth$^{\star}$ which are powers of two.
However, within the first 4 even harmonics, three are powers of 2: 2, 4 and 8. Hence, given the general tendency for upper harmonics to decrease in amplitude, the consonance
of the fifth$^{\star}$ is almost perfect, much more so than what occurs in the standard music system with equal intervals, the 12TET intonation. Table~\ref{table1}
summarises the elastic overtone system proposed, comparing the position of harmonics and the consonant qualities of the perfect fifth, the 12TET fifth and the fifth$^{\star}$,
up to the first 16 harmonics.






\begin{table}[!ht]
\begin{adjustwidth}{-2.6in}{0in} 
\centering
\caption{\textbf{Comparison of overtone frequencies.}}
\begin{tabular}{|l+l|l|l|l||l|l|l|l|}
\hline
\textbf{Harmonic} & \textbf{Fundamental}\hspace {-5pt} & \textbf{Octave}\hspace {-5pt} & \textbf{Perfect fifth}\hspace {-3pt}  \hspace {-6pt}  & \textbf{12 TET fifth}  & \textbf{Fundamental$^{\star}$}\hspace {-6pt}
& \textbf{Octave$^{\star}$} & \textbf{Fifth$^{\star}$} \\ \thickhline \hline
1 & 1 & 2 &  3/2      &  $2^{7/12}$                 & 1                            & $x$               & $x^{7/12}$                                                           \\ \hline
2 & 2 & 4 &  3,\qquad \quad 0     &  $2 \cdot 2^{7/12}$, 0.113  & $x$                          & $x^{2}$            & $x               \cdot x^{7/12}$, \qquad \qquad \quad 0   \\ \hline
3 & 3 & 6 &  9/2      &  $3 \cdot 2^{7/12}$         & $\frac{1}{2}(x+x^{2})$   &  $\frac{1}{2}(x^{2}+x^{3})$ & $\frac{1}{2}(x+x^{2})       \cdot x^{7/12}$                     \\ \hline
4 & 4 & 8 &  6,\qquad \quad 0     &  $4 \cdot 2^{7/12}$, 0.113  & $x^{2}$                       & $x^{3}$ & $x^{2}                      \cdot x^{7/12}$,\qquad \qquad \quad 0    \\ \hline
5 & 5 & 10&  15/2     &  $5 \cdot 2^{7/12}$         & $\frac{1}{4}(3x^{2}+x^{3})$    &  $\frac{1}{4}(3x^{3}+x^{4})$ & $\frac{1}{4}(3x^{2}+x^{3}) \cdot x^{7/12}$   \\ \hline
6 & 6 & 12&  9,\qquad \quad 0     &  $6 \cdot 2^{7/12}$, 0.113  & $\frac{1}{4}(2x^{2}+2x^{3})$   & $\frac{1}{4}(2x^{3}+2x^{4})$ & $\frac{1}{4}(2x^{2}+2x^{3}) \cdot x^{7/12}$, 0.15   \\ \hline
7 & 7 & 14&  21/2     &  $7 \cdot 2^{7/12}$         & $\frac{1}{4}(x^{2}+3x^{3})$    & $\frac{1}{4}(x^{3}+3x^{4})$  & $\frac{1}{4}(x^{2}+3x^{3}) \cdot x^{7/12}$              \\ \hline
8 & 8 & 16&  12,\qquad \, 0    &  $8 \cdot 2^{7/12}$, 0.113  & $x^{3}$                       & $x^{4}$ & $x^{3}                       \cdot x^{7/12}$,\qquad \qquad \quad 0   \\ \hline
9 & 9 & 18&  27/2     &  $9 \cdot 2^{7/12}$         & $\frac{1}{8}(7x^{3}+x^{4})$    & $\frac{1}{8}(7x^{4}+x^{5})$ & $\frac{1}{8}(7x^{3}+x^{4})  \cdot x^{7/12}$                \\ \hline
10 & 10 & 20& 15,\qquad \, 0    &  $10 \cdot 2^{7/12}$, 0.113 & $\frac{1}{8}(6x^{3}+2x^{4})$  & $\frac{1}{8}(6x^{4}+2x^{5})$ & $\frac{1}{8}(6x^{3}+2x^{4}) \cdot x^{7/12}$, 0.18  \\ \hline
11 & 11 & 22& 33/2     &  $11 \cdot 2^{7/12}$        & $\frac{1}{8}(5x^{3}+3x^{4})$  & $\frac{1}{8}(5x^{4}+3x^{5})$ & $\frac{1}{8}(5x^{3}+3x^{4}) \cdot x^{7/12}$            \\ \hline
12 & 12 & 24& 18,\qquad \, 0    &  $12 \cdot 2^{7/12}$, 0.113 & $\frac{1}{8}(4x^{3}+4x^{4})$  & $\frac{1}{8}(4x^{4}+4x^{5})$ & $\frac{1}{8}(4x^{3}+4x^{4}) \cdot x^{7/12}$, 0.15    \\ \hline
13 & 13 & 26& 39/2     &  $13 \cdot 2^{7/12}$        & $\frac{1}{8}(3x^{3}+5x^{4})$  & $\frac{1}{8}(3x^{4}+5x^{5})$ & $\frac{1}{8}(3x^{3}+5x^{4}) \cdot x^{7/12}$              \\ \hline
14 & 14 & 28& 21,\qquad \, 0    &  $14 \cdot 2^{7/12}$, 0.113 & $\frac{1}{8}(2x^{3}+6x^{4})$  & $\frac{1}{8}(2x^{4}+6x^{5})$ & $\frac{1}{8}(2x^{3}+6x^{4}) \cdot x^{7/12}$, 0.065  \\ \hline
15 & 15 & 30& 45/2     &  $15 \cdot 2^{7/12}$        & $\frac{1}{8}(x^{3}+7x^{4})$   & $\frac{1}{8}(x^{4}+7x^{5})$ & $\frac{1}{8}(x^{3}+7x^{4})   \cdot x^{7/12}$              \\ \hline
16 & 16 & 32& 24,\qquad \, 0    &  $16 \cdot 2^{7/12}$, 0.113 & $x^{4}$                      & $x^{5}$ & $x^{4}                       \cdot x^{7/12}$, \qquad \qquad \quad 0  \\ \hline

\end{tabular}
\begin{flushleft} The table shows the overtone structure of a fundamental note, its octave and its fifth in the two systems discussed in the text, for the first 16 harmonics.
  In the standard music system all overtones are integer multiples of the base frequency, taking the base frequency of the fundamental note as 1, all its overtones have a frequency
  which equal their corresponding harmonic number entry. For a perfect fifth, the base frequency is 3/2 of that of the fundamental note, and harmonics follow as integer multiples
  of 3/2. Notice that all even harmonics of the perfect fifth coincide with a harmonic of the fundamental having an entry number 3/2 higher, e.g. the frequency of the 2nd harmonic
  of the perfect fifth corresponds to the frequency of the 2$\times$ 3/2 = 3rd harmonic of the fundamental. The zeros after a comma following the frequency of the even harmonics
  for the perfect fifth denote the frequency percentage difference to the $ 3/2 \times$ harmonic of the fundamental, which in this case match exactly. For a 12TET tuning of the
  standard music system, the base frequency of the fifth is now given by $2^{7/12}$, and its overtones follow integer multiples of this value. No even harmonics of the 12TET fifth
  match the corresponding harmonic of the fundamental, all have an error of 0.113 percent, as shown. For the alternative music system the harmonics of the fundamental$^{\star}$
  appear at the values shown. The fifth$^{\star}$ has a base frequency of $x^{7/12}$ that of the base frequency of the fundamental, and harmonics which are multiples of this
  quantity times the frequencies of the harmonics of the fundamental. Now, harmonics with entries being powers of 2 are exact, and match the corresponding harmonics of the
  fundamental, as happens in the perfect fifth of the standard music system. All other even harmonics have a small error, comparable to what appears in the standard music system
  with 12TET intonation. Within the first 8 harmonics, where the perfect fifth has 4 exact matching overtones, the elastic overtone music system discussed has three such perfect
  matchings including the first two cases of the perfect fifth, while in the standard music system with 12TET intonation none remain. Notice that the structure of the octave is
  preserved in the octave$^{\star}$, in that all harmonics of the octave$^{\star}$ occur also in the fundamental$^{\star}$, at the same positions as in the octave and the fundamental
  of the standard music system.

\end{flushleft}
\label{table1}
\end{adjustwidth}
\end{table}

We see from  Table~\ref{table1} that the even harmonics$^{\star}$ of the fifth$^{\star}$ which do not exactly coincide with the corresponding perfect fifth harmonic of the
fundamental, are about as off from perfect as those in the 12TET fifth, which are all consistently off by the same \% in frequency with respect to the perfect fifth. The
12TET fifth yields no exact matching harmonics.

In going to the major third, an interval of four semitones, as we have seen, there, the most important consonance in the standard music system is that the fourth harmonic of the perfect
major third exactly matches the fifth harmonic of the fundamental. In the 12TET intonation, these two harmonics are out by the fractional difference between 5 and $4\times 2^{4/12}$,
0.79\% or 14 cents. In the Elastic Overtone proposal described, the corresponding offset is now the fractional difference between $0.25(3x^{2}+x^{3})$ and $x^{7/3}$, 0.88\% or 15 cents.
Thus, we see that in the proposed system which was calibrated to obtain an optimal consonance in the fifths, the major thirds are comparably as consonant as under the 12TET system.
Interestingly, in going to an interval of 5 semitones, the fourth, the third harmonic$^{\star}$ of the fourth$^{\star}$ in the Elastic Overtone proposal, exactly matches the fourth
harmonic$^{\star}$ of the fundamental$^{\star}$. Also as in the case of the perfect fourth, the 6th harmonic$^{\star}$ of the fourth$^{\star}$ exactly matches the 8th harmonic$^{\star}$ of the
fundamental$^{\star}$, consonances which are lost in 12TET. The fact that the fourth$^{\star}$ is also close to perfect leads to the expectation of a close correspondence between EO and JI.

\section*{Implementation}

We now present an implementation of the proposed music system, for a particular timbre choice. This example is largely arbitrary and is given simply
as a means to give a concrete example, amongst the many parameter choices at this point. First, we choose a set of amplitudes for the harmonics of
a 12TET system. The series of amplitudes for harmonics which define this timbre is given in Table~\ref{table3}, which also gives the explicit values of the frequencies
of the first 16 harmonics in both systems, in accordance with the model presented in the previous section, and summarised in Table~\ref{table1}. This is used to produce
a 5 octave 12TET standard music system, all semitones from A0 to A5, with A4 = 440 Hz. The fundamental frequency of any other note $n$ semitones away is given by
$220 \times 2^{n/12}$. The timbre is kept constant across all 4 octaves, although this is of course just a simplification for the example being presented, in acoustic
instruments the timbre structure of a note varies with the fundamental frequency of it.

To now construct the modified music system, we choose an equivalence point where the fundamental frequency of
a given note will be the same in both systems. In this particular implementation we pick A3=220 Hz. as the equivalence point. This has the advantage of being an exactly defined
integer value, and is approximately in the middle of the entire frequency range of the chosen 12TET system, which will reduce the maximum frequency offset between both systems
at any given note. Thus, the fundamental frequency of A3$^{\star}$ will also be 220 Hz. The harmonics of this note will have the same timbre structure as those of the 12TET system,
amplitudes given also by the same values of Table~\ref{table3}. However, the position in frequency of the harmonics$^{\star}$ of A3$^{\star}$ will now be fixed by the formulas
given in Table~\ref{table1}. Since the harmonics$^{\star}$ of any note all occur at slightly lower frequencies in the modified music system with regards to the standard 12TET
system, given the timbre chosen, with substantial power across many harmonics, even though the fundamental frequency of A3$^{\star}$ is equal to that of A3, the former sounds
slightly flat with respect to the latter. To complete the modified music system, the fundamental frequency of any other note $n$ semitones away will be given by $220 \times
x^{n/12}$, and its harmonics$^{\star}$ will follow the structure described for A3$^{\star}$.

All 12TET and Elastic Overtone sounds were uploaded to a Kontakt instrument (sampling software). A simple ADSR envelope and Pitch Bend modulation, as well as a mild reverb,
were identically applied to both music systems. A special function was included in order to easily switch between both systems within the same instrument. Finally, the .niki
file produced by Kontakt was uploaded as a plugin inside a ProTools DAW through which all the following examples were programmed via MIDI.

\begin{table}[h!]
\caption{Timbre for implementation given.}
\label{table3}
\begin{tabular}{llll}
\textbf{Harmonic} &\textbf{Frequency ratio 12TET} & \textbf{Frequency ratio EO} &\textbf{Amplitude}  \\
      1  & 1  & 1       & 1.0     \\
      2  & 2  & 1.9726  & 0.95    \\
      3  & 3  & 2.9318  & 0.93    \\
      4  & 4  & 3.8911  & 0.91    \\
      5  & 5  & 4.8372  & 0.89    \\
      6  & 6  & 5.7833  & 0.87    \\
      7  & 7  & 6.7293  & 0.7     \\
      8  & 8  & 7.6754  & 0.6     \\    
      9  & 9  & 8.6086  & 0.5     \\
      10 & 10 & 9.5417  & 0.4     \\
      11 & 11 & 10.4748 & 0.3     \\
      12 & 12 & 11.4079 & 0.2     \\
      13 & 13 & 12.3410 & 0.1     \\
      14 & 14 & 13.2742 & 0.05    \\
      15 & 15 & 14.2073 & 0.025   \\
      16 & 16 & 15.1404 & 0.0125  \\
      
\end{tabular}
\begin{flushleft} The second and third columns give the frequency ratio of the corresponding harmonics with respect to the frequency of the fundamental for the
  two systems discussed. The fourth column gives the relative amplitudes with respect to that of the amplitude of the fundamental frequency for the timber structure of both the
  12TET and elastic overtone examples implemented, both kept fixed for all notes modelled.
\end{flushleft}
\end{table}

Having a standard 12TET and a modified music system, we can now compare the sound of each. To do so, care was taken to generate all the notes in both systems using
identical timbres, as shown in Table~\ref{table3}. We encourage the reader to take the time to listen to the sound files provided with headphones, and to listen repeatedly.
For all sound files presented, the first in a series gives the 12TET system, and the second presents the same notes, in the modified elastic overtone system (EO). The scores
for these first 4 comparative examples are shown in Figures 1-4.

The first sound file shows the same chromatic scale in the 12TET standard system and in the modified EO system, ranging from A3 to A4. Note that in the first three
audio examples, the equivalence note of A3=A3$^{\star}$ is contained in the notes played. It should be noted that upon repeated listening of the chromatic scale in the
EO system, a relative smoothness becomes apparent when compared to the equivalent scale in 12-TET, asides from a very marginal frequency offset, the octave$^{\star}$ being
slightly shorter than the octave.

Sound file 2 gives a series of parallel fifths played in a C major scale for both music systems, presented in the same order as the previous example. In this case, the
smoothness becomes much more readily apparent due to the higher consonance of harmonics, even for the diminished fifth at B3. Indeed, there is none of the ``raspiness"
one has grown accustomed to in 12-TET fifths; listeners who have worked with Just Intonation (JI) would find great familiarity of character between JI fifths and those
of the EO system since they are both intervals in which the third harmonic of the fundamental exactly matches the second one of the fifth, even if the exact frequencies
of the notes in JI and the EO system are not equal.

Sound file 3 is an example of parallel triads played in C major in both tuning systems, and sound file 4 is an example of parallel tritones, where the notes
cannot be within the scale of C Major. In sound file 3, we draw especial attention to the seventh triad, which is diminished, as an interesting point of comparison between
the two tuning systems; the Locrian mode is usually shunned or at the very least treated with some trepidation, but it sounds different in the EO system. Again, an
immediate contrast in smoothness becomes apparent, and initial discomfort for professional listeners trained in 12-TET will vanish after repeated listening.

The real test of the tuning system, however, comes when it is applied to actual music. To this end, we present 2 examples of music with increasing harmonic complexity. Both
examples were transformed into a MIDI file and performed by a sequencer to ensure the execution of the pieces is absolutely identical in each tuning system. We begin with
the Goldberg Variations BWV 988 variation no. 1 by Johann Sebastian Bach, we follow that with Fantasie Impromptu in C-sharp minor and by Fr\'{e}d\'{e}ric Chopin.
This exercise really helps to highlight the powerful differences between 12TET and EO. Again, repeated listening of both tuning systems will reveal the significant perceptual
differences when music is played in both tuning systems. The differences become more acute with increasing harmonic complexity. We encourage repeated listening to appreciate the
interesting nuances of EO when compared to 12TET.

\begin{figure}[!h]
   \includegraphics[height=1.3cm,width=14.0cm]{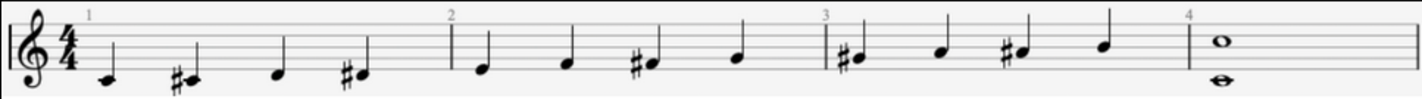}
\caption{\textbf{Score for sound File 1.}
This Figure shows the score for sound file 1, a chromatic scale, which is played in the standard 12TET system first and in the proposed elastic overtone system after.}
\label{fig1}
\end{figure}

\begin{figure}[!h]
   \includegraphics[height=3.0cm,width=8.8cm]{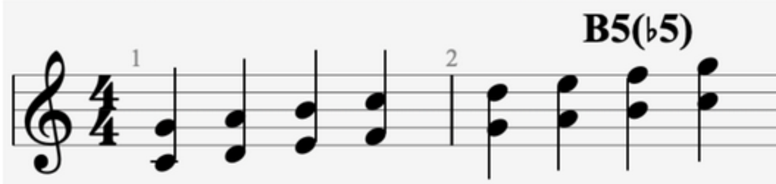}
\caption{\textbf{Score for sound File 2.}
This Figure shows the score for sound file 2, a series of parallel fifths, which are played in the standard 12TET system first and in the proposed elastic overtone system after.}
\label{fig1}
\end{figure}

\begin{figure}[!h]
  \includegraphics[height=3.0cm,width=12.0cm]{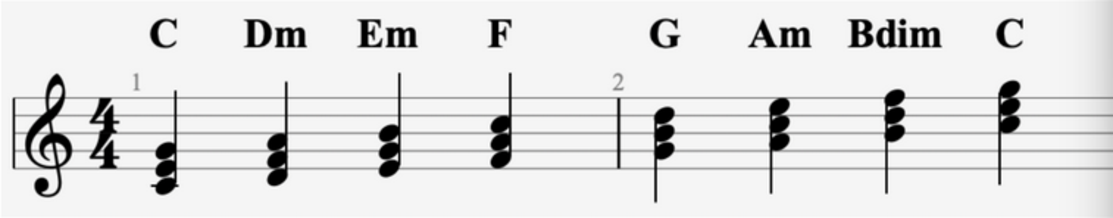}
\caption{\textbf{Score for sound File 3.}
This Figure shows the score for sound file 3, a series of chords, which are played in the standard 12TET system and the proposed modified system, respectively.}
\label{fig1}
\end{figure}

\begin{figure}[!h]
  \includegraphics[height=3.0cm,width=12.0cm]{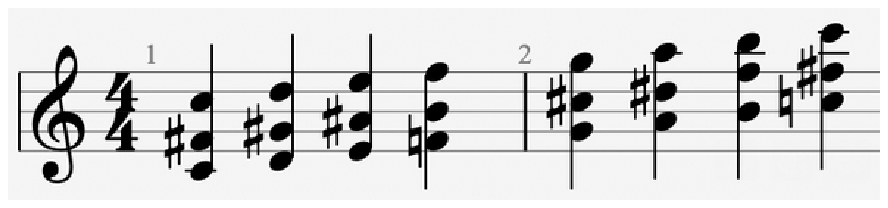}
\caption{\textbf{Score for sound File 4.}
This Figure shows the score for sound file 4, a series of tritones, which are played in the standard 12TET system and the proposed modified system, respectively.}
\label{fig1}
\end{figure}




\section*{Conclusion}

We have presented a modified music system which is analogous to the standard one and shares many features of the standard 12TET tuning scheme. A primary
octave$^{\star}$ division is preserved such that notes at the same position within a different octave$^{\star}$ sound the same, for the same reason of sharing
harmonics$^{\star}$ as happens in the standard music system. The octave$^{\star}$ interval is no longer a factor of 2 in frequency, but of $x=1.9725786270914...$.
Necessarily, harmonics$^{\star}$ are no longer integer multiples of the fundamental frequency of any given note, but arranged so that the octave
structure of the standard music system is preserved. Interestingly, most of the resonances of the perfect fifth and perfect fourth of the standard music system
under Just Intonation are retained at the lower 8 harmonics$^{\star}$ by construction, this time in a fully transposable system where semitones are equally
spaced within the octave$^{\star}$.

We present a comparison of the 12TET intonation of the standard music system to the proposed elastic overtone system, for a particular arbitrary
timber selection, and a couple of well known musical pieces. The added consonance of the modified system presents the opportunity of interesting
musical innovations not previously explored and results in a musical experience highly reminiscent of Just Intonation, at the expense of a slightly
shortened octave.

\section*{Supporting information}

\paragraph*{S1 Sound File.}
\label{S1_SoundFile}
\textbf{Chromatic Scales.} Chromatic.mp3, a sound file comparing a chromatic scale in the standard 12TET (appearing first) and in the elastic overtone systems, last.

\paragraph*{S2 Sound File.}
\label{S2_SoundFile}
\textbf{Parallel Fifths.} ParallelFifthsMajorScale.mp3, a sound file comparing a series of parallel fifths in a major scale in the standard 12TET (appearing first)
and the elastic overtone systems, last.

\paragraph*{S3 Sound File.}
\label{S3_SoundFile}
\textbf{Chords1.} ParallelTriadsMajorScale.mp3, a sound file comparing a series of parallel triads in a major scale in the standard 12TET (appearing first) and in the elastic
overtone systems, last.

\paragraph*{S4 Sound File.}
\label{S3_SoundFile}
\textbf{Chords2.} ParallelTritonesMajorScale.mp3, a sound file comparing a series of parallel tritones in a major scale in the standard 12TET (appearing first) and in the elastic
overtone systems, last.

\paragraph*{S5 Sound File.}
\label{S1_Video}
\textbf{BWV 988.} Bach.mp3, a sound file comparing Variation 1 of Bach's Goldberg Variations in the standard 12TET (appearing first) and in the elastic overtone systems, last.

\paragraph*{S6 Sound File.} 
\label{S1_Appendix}
\textbf{Fantasie-Impromptu in C sharp Minor Op.66.}  Chopin.mp3, a sound file comparing the Chopin's Fantasie-Impromptu in C sharp Minor Op.66 in the standard 12TET
(appearing first) and in the elastic overtone systems, last.


\section*{Acknowledgments} Luis Nasser gratefully acknowledges the support from the NSF award PHY - 2110425. Xavier Hernandez acknowledges financial
assistance from SNII-SECIHTI and UNAM DGAPA grant IN-102624. Pablo Garc\'{\i}a-Valenzuela acknowledges the full support from Sistema Nacional de
Creadores de Arte awards SNCA-11/15/20. M\'exico.

\nolinenumbers

%
%

\pagebreak

\end{document}